\documentclass[10pt,twoside,english]{elsarticle}
\usepackage[T1]{fontenc}
\usepackage[latin9]{inputenc}
\usepackage{geometry}
\usepackage{amsmath}
\usepackage{graphicx}
\usepackage{setspace}
\usepackage{esint}


\usepackage{babel}
\begin{document}

\title{\textbf{Qualitative modeling of the dynamics of detonations with
losses}}

\author{Luiz M. Faria and Aslan R. Kasimov}

\address{Applied Mathematics and Computational Science\\
Room 4-2226, 4700 King Abdullah University of Science and Technology\\
Thuwal 23955-6900, Saudi Arabia}

\ead{aslan.kasimov@kaust.edu.sa}
\begin{abstract}
We consider a simplified model for the dynamics of one-dimensional
detonations with generic losses. It consists of a single partial differential
equation that reproduces, at a qualitative level, the essential properties
of unsteady detonation waves, including pulsating and chaotic solutions.
In particular, we investigate the effects of shock curvature and friction
losses on detonation dynamics. To calculate steady-state solutions,
a novel approach to solving the detonation eigenvalue problem is introduced
that avoids the well-known numerical difficulties associated with
the presence of a sonic point. By using unsteady numerical simulations
of the simplified model, we also explore the nonlinear stability of
steady-state or quasi-steady solutions.\end{abstract}
\maketitle

\section{Introduction}

A gaseous detonation is a phenomenon exhibiting rich dynamical features.
One-dimensional planar detonations propagate with a velocity that
can be steady, periodic, or chaotic \citep{Ng2005}. In multiple dimensions,
the detonation front includes complex structures resulting in cellular
patterns formed by triple-point trajectories \citep{Lee-2008,fickett2011detonation}.
Quasi-steady curved detonations, characteristic of condensed explosives,
possess multiple-valued solutions at a given curvature \citep{bdzil2012theory}.
The same multiplicity of solutions exists in one-dimensional detonations
in the presence of heat and momentum losses \citep{Zeldovich1940,zeldovich1960theory,brailovsky2012combustion,Semenko-2013}.
This range of complex dynamical properties of detonations poses a
challenge in terms of theoretical understanding of conditions in which
they arise and of features they exhibit. The linear stability theory
for idealized systems, asymptotic theories of weakly curved detonation,
and other asymptotic models have significantly advanced our understanding
of the detonation phenomenon (see recent reviews in \citep{zhang2012shock}).
However, many problems still require further theoretical study, including
the mechanism of detonation cell formation, the nature of critical
conditions of detonation propagation in systems with losses, the linear
and nonlinear instability in systems described by complex reactions
and equations of state, and others. 

Elucidation of key physical mechanisms of the complex phenomena of
detonation dynamics is greatly facilitated by simplified models, including
those of \emph{ad hoc} nature \citep{Fickett:1979ys}. Such models
can highlight in the clearest possible way the processes responsible
for a particular qualitative trait in the observed dynamics. A wide
range of dynamical properties of one-dimensional detonations, including
chaotic solutions, is reproduced in \citep{Radulescu:2011fk,tang2012dynamics}
with a simple extension of Fickett's analog \citep{Fickett:1979ys}
to model the chemical reaction with a well-defined induction zone
followed by a heat-release zone. In \citep{kasimov2013model,FariaKasimovRosales-SIAM2014},
it was shown that a model consisting of just a single scalar equation
is also capable of qualitatively capturing the dynamics of one-dimensional
detonations in the reactive Euler equations, including instability
and chaos. The most important implication of these simplified models
is that the true nature of the complex dynamics of detonations appears
to be governed by a simple mechanism, thus providing a strong indication
that a rational reduction of the reactive Euler equations that retains
the same essential physical ingredients as the simple models may be
feasible.

The model in \citep{kasimov2013model} is given by the following equation:
\begin{equation}
\frac{\partial u}{\partial t}+\frac{\partial}{\partial x}\left(\frac{u^{2}}{2}-\frac{uu_{s}}{2}\right)=f\left(x,u_{s}\right),\label{eq:KFR-planar}
\end{equation}
where $x\le0$ is the reaction zone behind the shock propagating from
left to right. Equation (\ref{eq:KFR-planar}) is written in a shock-attached
frame; the shock location is hence at $x=0$ at all times, $t$. The
unknown, $u\left(x,t\right)$, plays the role of, e.g., pressure,
while $u_{s}$ is the solution $u$ evaluated at the shock, and it
is related to the shock speed through shock conditions. The forcing
function, $f$, is chosen to mimic the behavior of the reaction rate
in the reactive Euler equations. In particular, it is taken to have
a maximum at some distance away from the shock, $x_{f}=x_{f}\left(u_{s}\right)$,
with function $x_{f}$ chosen to depend sensitively on the shock state,
$u_{s}$. The following choice, 
\begin{equation}
f=\frac{a}{\sqrt{4\pi\beta}}\exp\left[-\frac{\left(x+u_{s}^{-\alpha}\right)^{2}}{4\beta}\right],\label{eq:f}
\end{equation}
where $a=\left[4\left(1+\mathrm{erf}\left(u_{s}^{-\alpha}/2\sqrt{\beta}\right)\right)\right]^{-1}$,
is used in numerical calculations below, as in \citep{FariaKasimovRosales-SIAM2014}.
In this form, the model is dimensionless with $u$ scaled so that
$u_{s}=1$ in the steady state. Parameters $\alpha$ and $\beta$
are analogous to the activation energy in the reactive Euler equations
with Arrhenius kinetics and to the ratio of the reaction-zone length
to the induction-zone length, respectively. Note that the total chemical
energy released corresponds to $\int_{-\infty}^{0}f(x,u_{s}(t))dx$,
which is constant for the forcing term (\ref{eq:f}) regardless of
the value of $u_{s}(t)$. This follows from $f\sim\lambda_{x}$, as
discussed in \citep{kasimov2013model,FariaKasimovRosales-SIAM2014}.
Thus, the total energy released is always the same even in the presence
of instabilities. 

Equation (\ref{eq:KFR-planar}) can be shown to be closely related
to the asymptotic model \citep{RosalesMajda:1983ly} derived from
the reactive Euler equations. From a physical point of view, an important
ingredient of the model is that it represents the nonlinear interaction
of two wave families: one moving slowly toward the shock and one moving
infinitely quickly away from the shock. The former is simply the wave
evolving along the Burgers characteristic. The wave moving infinitely
fast is implied by the presence of the shock state, $u_{s}$, directly
in (\ref{eq:KFR-planar}), such that the solution, $u\left(x,t\right)$,
at any given time, $t$, at any location, $x$, depends on the shock
state at that particular time. This non-locality is a result of taking
to an extreme the asymptotic idea that the waves reflecting from the
shock propagate much faster than the waves moving toward the shock
from the reaction zone. Another element of the model that is of physical
significance is that when $f$ has a maximum at some distance away
from the shock, and the position of this maximum depends sensitively
on the shock state, the system represents a kind of a resonator that
amplifies the waves moving back and forth between the shock and the
region around $x_{f}$. This resonant amplification is a real mechanism
for instability as observed in the simulations of pulsating solutions
of (\ref{eq:KFR-planar}) \citep{FariaKasimovRosales-SIAM2014}.

\section{A model with generic losses}

Our focus here is to explore the effect of generic losses on the solutions
of (\ref{eq:KFR-planar}). For this purpose, we modify the forcing
in (\ref{eq:KFR-planar}) to add a damping term, 
\begin{equation}
\frac{\partial u}{\partial t}+\frac{\partial}{\partial x}\left(\frac{1}{2}u^{2}-Du\right)=f\left(x,D\right)-g(x,u,\varphi).\label{eq:KFR-damped}
\end{equation}
Here, $D=u_{s}/2$ is the detonation speed, which is obtained using
the Rankine-Hugoniot conditions with the state upstream of the shock
taken to be $u=0$ \citep{FariaKasimovRosales-SIAM2014}, $\varphi$
is a parameter of the problem, which may be time dependent, and $g$
is a function that represents the loss. Friction losses are modeled
by taking $g=c_{f}u|u|$, with the friction coefficient $c_{f}$,
while the effects of curvature are modeled by taking $g=\kappa u^{2}/\left(1+\kappa x\right)$,
where $\kappa$ is the shock curvature, generally dependent on time.

\subsection{Steady and quasi-steady solutions}

If $\varphi$ is a constant, then we can find steady-state solutions
of (\ref{eq:KFR-damped}). If $\varphi$ is time-dependent, but slowly
varying in time, then we can find quasi-steady solutions of (\ref{eq:KFR-damped}).
In both cases, the problem requires solving the ordinary differential
equation (ODE),
\begin{equation}
\left(u-D\right)u'=f\left(x,D\right)-g(x,u,\varphi),\label{eq:quasi-steady-ode}
\end{equation}
on $x\in\left[a,0\right]$ with $u(0)=2D$ as the shock condition.
Here and below, primes denote the derivative, $d/dx$. The left end
of the integration region is either $a=-\infty$ or the sonic locus,
$a=x_{*}$, where $u-D=0$. The main problem is to determine the detonation
speed, $D$, such that the corresponding solution, $u\left(x,t\right)$,
of (\ref{eq:quasi-steady-ode}) is a smooth function of $x$. This
is a nonlinear eigenvalue problem for $D$ because such smooth solutions
do not necessarily exist for every $D$ at a given $\varphi$. For
physically interesting choices of $f$ and $g$, there usually exists
a sonic point where $u=D$, which is a singular point of (\ref{eq:quasi-steady-ode}).
For smoothness of $u$, it is necessary that the right-hand side of
(\ref{eq:quasi-steady-ode}) vanishes at the sonic point. These conditions
constitute the generalized Chapman-Jouguet conditions of detonation
theory and serve to determine the eigenvalue relation, $H(D,\varphi)=0$,
that yields $D$ for a given $\varphi$. Typically, $D\left(\varphi\right)$
is a multiple-valued function having a turning-point shape.

The nonlinear ODE (\ref{eq:quasi-steady-ode}) cannot, in general,
be solved analytically. Therefore, a numerical integration method
is required. In one such method, for a trial value of $D$, (\ref{eq:quasi-steady-ode})
is integrated from $x=0$ toward $x=a$. The correct value of $D$
has to correspond to $u-D=0$ and $f-g=0$ at $x=x_{*}$. These conditions
are not satisfied in most cases, and, therefore, equation (\ref{eq:quasi-steady-ode})
is very stiff as $u\rightarrow D$, making the numerical integration
prohibitively expensive and/or inaccurate. As an alternative to this
method, the sonic locus, $x_{*}(D,\varphi)$, is found first for a
trial value of $D$. Then, the solution of (\ref{eq:quasi-steady-ode})
is found analytically in the neighborhood of $x_{*}$ in order to
get out of the sonic point by a small step to $x_{*}+\Delta x$, with
a subsequent numerical integration from $x_{*}+\Delta x$ toward the
shock. For the correct value of $D$, the Rankine-Hugoniot conditions
at $x=0$ must be satisfied. This algorithm is more robust numerically.
However, its drawback is that it requires the knowledge of the sonic
state and the ability to solve the equation (or the system of equations,
in general) in the neighborhood of the sonic locus analytically. Even
though, in our case, it is straightforward to do so, in more complicated
problems, this approach is not feasible \citep{Semenko-2013}.

Here, we propose a different algorithm that is much simpler, more
robust, and easier to generalize (see Appendix for the general version
of the algorithm). The key idea of the method is a change of the dependent
variable that eliminates the singularity from the governing ODE. Specifically,
we introduce $z=\left(u-D\right)^{2}$ as a new variable instead of
$u$. Then, (\ref{eq:quasi-steady-ode}) becomes 
\begin{equation}
z'=2\left(f\left(x,D\right)-g(x,u,\varphi)\right),\label{eq:z-eqn}
\end{equation}
which has a regular right-hand side. Notice that the inverse of the
transformation from $u$ to $z$ is double-valued, $u=D\pm\sqrt{z}$.
At the shock, $u(0)=2D>D$, and, therefore, between the shock and
the sonic point, we have $u=D+\sqrt{z}$. Hence 
\begin{equation}
z'=2\left(f\left(x,D\right)-g(x,D+\sqrt{z},\varphi)\right).\label{eq:z-eqn-2}
\end{equation}
Downstream of the sonic point, the square root changes its branch.
Therefore, $u=D-\sqrt{z}$. The sonic condition in the new variable
is very simple: $z'=0$ at $z=0$. These conditions are clearly independent
of the specific form of the right-hand side of (\ref{eq:quasi-steady-ode}).
The main advantages of the new algorithm are that the equations are
no longer stiff and that the sonic conditions are very simple. If
the solution beyond the sonic point is required, then $z'=2\left(f\left(x,D\right)-g(x,D-\sqrt{z},\varphi)\right)$
must be solved at $x<x_{*}$. 

The substitution employed here is applicable to a wide range of problems
\citep{Semenko-2013}. For example, the problem of finding a quasi-steady
solution of a curved expanding detonation leads to the ODE for the
flow velocity (e.g., \citep{bdzil2012theory}): 
\begin{equation}
\frac{du}{d\lambda}=\frac{\Phi}{u^{2}-c^{2}}\frac{u}{\omega},
\end{equation}
where $\omega=k\left(1-\lambda\right)\exp(-\gamma\vartheta/c^{2})$
is the reaction rate, $\vartheta$ is the activation energy, $\Phi=\left(\gamma-1\right)q\omega-\kappa c^{2}\left(u+D\right)$,
$\kappa$ is the shock curvature, $q$ is the heat release, and $c^{2}=\gamma p_{0}+\left(\gamma-1\right)\left[\left(D^{2}-u^{2}\right)/2+q\lambda\right]$.
The integration domain is $0\leq\lambda\leq1$ with $u(0)=u_{s}(D)$
given by the Rankine-Hugoniot condition\@. The sonic singularity
here occurs at $u=c$ and hence we introduce $z=\left(u-c\right)^{2}$
to obtain
\begin{equation}
\frac{dz}{d\lambda}=2\left(u-c\right)\left(1-\frac{\partial c}{\partial u}\right)\frac{du}{d\lambda}=2\left(1-\frac{\partial c}{\partial u}\right)\frac{\Phi u}{\omega},
\end{equation}
which is regular at the sonic point. After the correct branch of the
inversion is obtained, the generalized Chapman-Jouguet condition at
the sonic point in terms of the new variables is that $dz/d\lambda=0$
at the sonic point, $\lambda=\lambda^{*}$, where $z(\lambda^{*})=0$.
This provides a much simpler and faster way of solving the generalized
Chapman-Jouguet condition and allows for integration from the shock
toward the sonic point without any difficulty.

\subsection{On linear stability analysis}

Once the steady or quasi-steady solutions are obtained, the question
of their linear stability arises. The problem without losses is analyzed
extensively in \citep{FariaKasimovRosales-SIAM2014}, where it is
shown that the analysis parallels that of the reactive Euler equations. 

We begin with the stability of steady-state solutions. Let $u_{0}(x)$
be the solution of 
\begin{equation}
\frac{d}{dx}\left(\frac{1}{2}u_{0}^{2}-D_{0}u_{0}\right)=f\left(x,D_{0}\right)-g(x,u_{0},\varphi),
\end{equation}
where $\varphi$ is a constant and $D_{0}$ is such that the generalized
Chapman-Jouguet condition is satisfied. Consider then a perturbation
of this solution of the form $D=D_{0}+\epsilon\sigma\exp(\sigma t)$
and $u=u_{0}(x)+\epsilon u_{1}(x)\exp(\sigma t)$, where $\sigma$
is the growth rate to be found. Inserting these expansions into (\ref{eq:KFR-damped})
yields
\begin{eqnarray}
\sigma u_{1}+\left(u_{0}u_{1}-D_{0}u_{1}-\sigma u_{0}\right)' & = & \sigma\frac{\partial f}{\partial D}\left(x,D_{0}\right)-u_{1}\frac{\partial g}{\partial u}(x,u_{0},\varphi),\label{eq:spec-stab-ode-steady}
\end{eqnarray}
which can be solved exactly to yield the eigenfunction, 
\begin{alignat*}{1}
u_{1}\left(x\right)=\frac{\sigma}{c_{0}(x)}e^{p(x,\sigma)} & \left[\int_{0}^{x}\left(\frac{\partial f}{\partial D}\left(x,D_{0}\right)+u_{0}'\right)e^{-p(\xi,\sigma)}d\xi+2D_{0}\right],
\end{alignat*}
where $c_{0}=u_{0}-D_{0}$ and 
\[
p(x,\sigma)=\int_{x}^{0}\left[\sigma+\frac{\partial g}{\partial u}(\xi,u_{0}(\xi),\varphi)\right]\frac{d\xi}{c_{0}(\xi)}
\]
are functions of $f$ and $g$, which are known in terms the steady-state
solution, $u_{0}\left(x\right)$. Requiring boundedness of the eigenfunctions
gives the dispersion relation 
\begin{equation}
\int_{x_{*}}^{0}\left(\frac{\partial f}{\partial D}\left(\xi,D_{0}\right)+u_{0}'\right)e^{-p(\xi,\sigma)}d\xi-2D_{0}=0,\label{eq:dispersion-relation}
\end{equation}
which is the same as in the ideal case with the only change due to
$g$ appearing in the expression for $p$. Hence, the stability analysis
of the equation with losses is very similar to the ideal case analyzed
in \citep{FariaKasimovRosales-SIAM2014}. 

For quasi-steady problems, the stability analysis is a bit subtler.
Consider 
\begin{equation}
\frac{\partial u}{\partial t}+\frac{\partial}{\partial x}\left(\frac{1}{2}u^{2}-Du\right)=f\left(x,D\right)-g(x,u,\varphi),
\end{equation}
where $\varphi$ is a slowly varying function of time. Then, the steady-state
solution for $u$ does not exist in general. We then consider solutions
that are slowly evolving in time by considering a slow time variable,
$\tau=\delta t$, $0<\delta\ll1$, such that $\varphi=\varphi\left(\tau\right)$.
Then,
\begin{equation}
\delta\frac{\partial u}{\partial\tau}+\frac{\partial}{\partial x}\left(\frac{1}{2}u^{2}-Du\right)=f\left(x,D\right)-g(x,u,\varphi(\tau)).\label{eq:slow-time-pde}
\end{equation}
Let $u_{\delta}(x,\tau)$ be the exact solution of (\ref{eq:slow-time-pde})
with $D=D_{\delta}(\tau)$ as the speed. Then, the spectral stability
of this solution requires looking at the evolution of $D=D_{\delta}(\tau)+\epsilon\sigma\exp(\sigma t)$
and $u=u_{\delta}(x,\tau)+\epsilon u_{\delta1}(x,\tau)\exp(\sigma t)$.
It is important to observe that these expansions express $O\left(1\right)$
time-scale variations around the slow, $O\left(1/\delta\right)$,
time-scale leading solution. Putting these expressions into (\ref{eq:slow-time-pde}),
we obtain, to first order, 
\begin{eqnarray}
\delta\frac{\partial u_{\delta1}}{\partial\tau}+\sigma u_{\delta1}+\frac{\partial}{\partial x}\left(u_{\delta}u_{\delta1}-D_{\delta}u_{\delta1}-\sigma u_{\delta}\right) & =\\
\sigma\frac{\partial f}{\partial D}\left(x,D_{\delta}\right)-u_{\delta1}\frac{\partial g}{\partial u}(x,u_{\delta},\varphi).
\end{eqnarray}
Next, we perform an asymptotic expansion in $\delta$: $u_{\delta}=u_{0}+O(\delta)$,
$u_{\delta1}=u_{1}+O(\delta)$, $D_{\delta}=D_{0}+O(\delta)$. Then,
to leading order, the quasi-steady solution satisfies 
\begin{equation}
\frac{d}{dx}\left(\frac{1}{2}u_{0}^{2}-D_{0}u_{0}\right)=f\left(x,D_{0}\right)-g(x,u_{0},\varphi),
\end{equation}
which, together with the shock and sonic conditions, gives the eigenvalue
problem for $D_{0}$. The linear stability equation is, to leading
order in $\delta$, given by the same equation as (\ref{eq:spec-stab-ode-steady})
and hence the dispersion relation is also given by (\ref{eq:dispersion-relation}).
Notice here that the implicit assumption $\partial u_{0}/\partial\tau=O(1)$
is required for the validity of the asymptotic expansion in $\delta$.
This is seen to break down at a turning point of the $D_{0}-\varphi$
curve if such a point exists.

\section{Numerical results}

\begin{figure}
\noindent \begin{centering}
\includegraphics[width=3in]{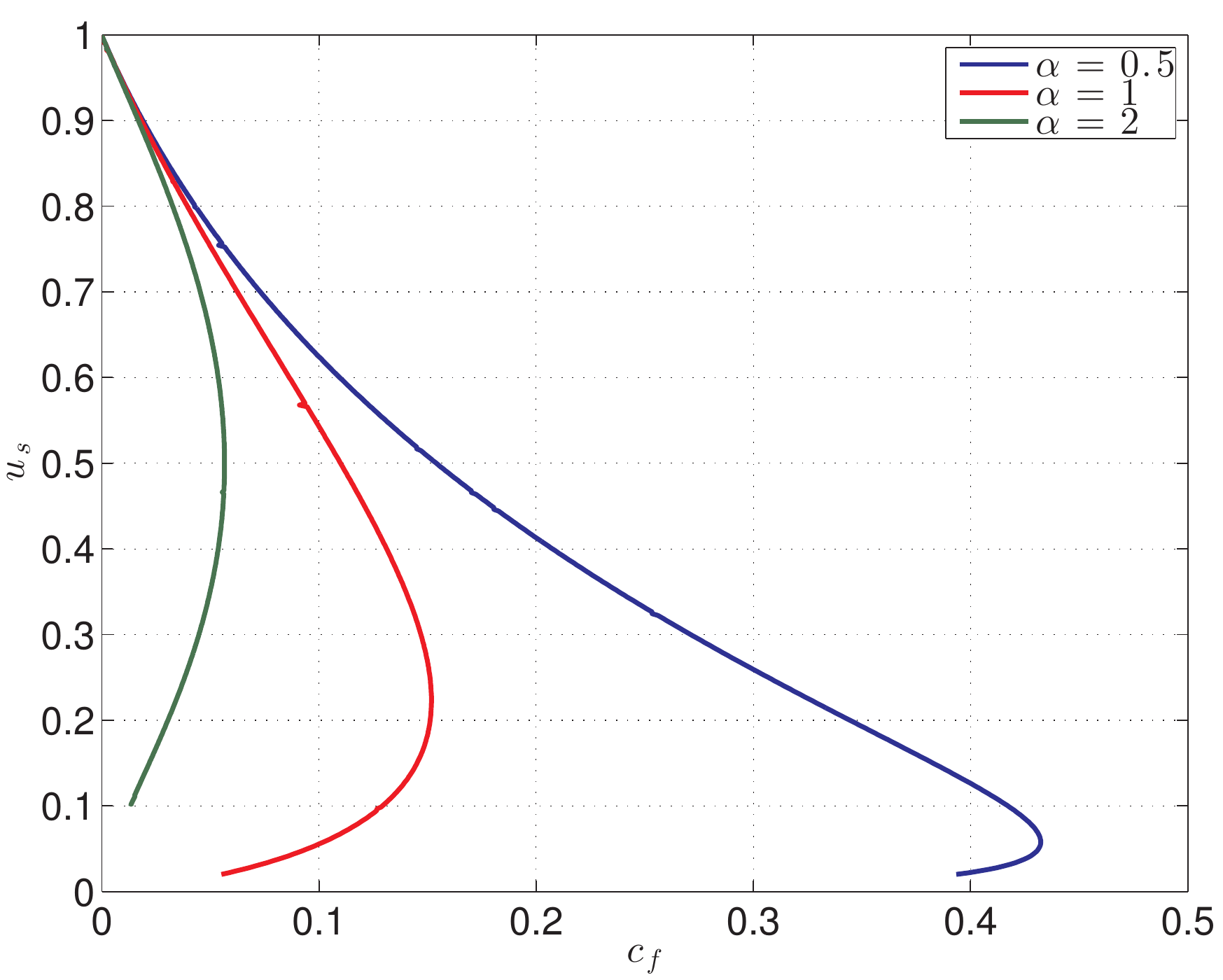}
\par\end{centering}

\caption{\label{fig:Steady-solution-with-friction} The $u_{s}-c_{f}$ relation
for the steady-state solution of (\ref{eq:model-with-losses}) for
detonation with friction.}
\end{figure}

In this section, we investigate numerically two types of losses, frictional
and those due to shock curvature. For detonation with frictional losses,
we consider 
\begin{equation}
\frac{\partial u}{\partial t}+\frac{\partial}{\partial x}\left(\frac{1}{2}u^{2}-Du\right)=f\left(x,D\right)-c_{f}u|u|,\label{eq:model-with-losses}
\end{equation}
where $x\in(-\infty,0]$ and $c_{f}$ is a constant friction coefficient.
The goal of the following calculations is to determine the role of
$c_{f}$ in the existence and structure of the steady-state solutions
of (\ref{eq:model-with-losses}). Figure \ref{fig:Steady-solution-with-friction}
shows the computed dependence of $u_{s}=2D$ on $c_{f}$, where we
can see the characteristic turning-point behavior with two solutions
coexisting at $c_{f}<c_{fc}$ and steady-state solutions no longer
existing if $c_{f}>c_{fc}$. 

Of particular interest is the question of stability of these steady-state
solutions. It is generally believed that the lower branch of the steady-state
$u_{s}$-$c_{f}$ curve is always unstable while the top branch can
be stable or unstable. In order to explore the nature of these instabilities,
we solve (\ref{eq:model-with-losses}) numerically using a second-order
finite volume Godunov's method with a min-mod limiter \citep{leveque2002finite}.
We begin with a perturbation around the steady-state solutions at
different locations of the $u_{s}$-$c_{f}$ curve, both on the top
and bottom branches. We choose $\alpha$ and $\beta$ such that the
corresponding ideal solution is stable. 

\begin{figure}
\noindent \begin{centering}
\includegraphics[height=2.4in]{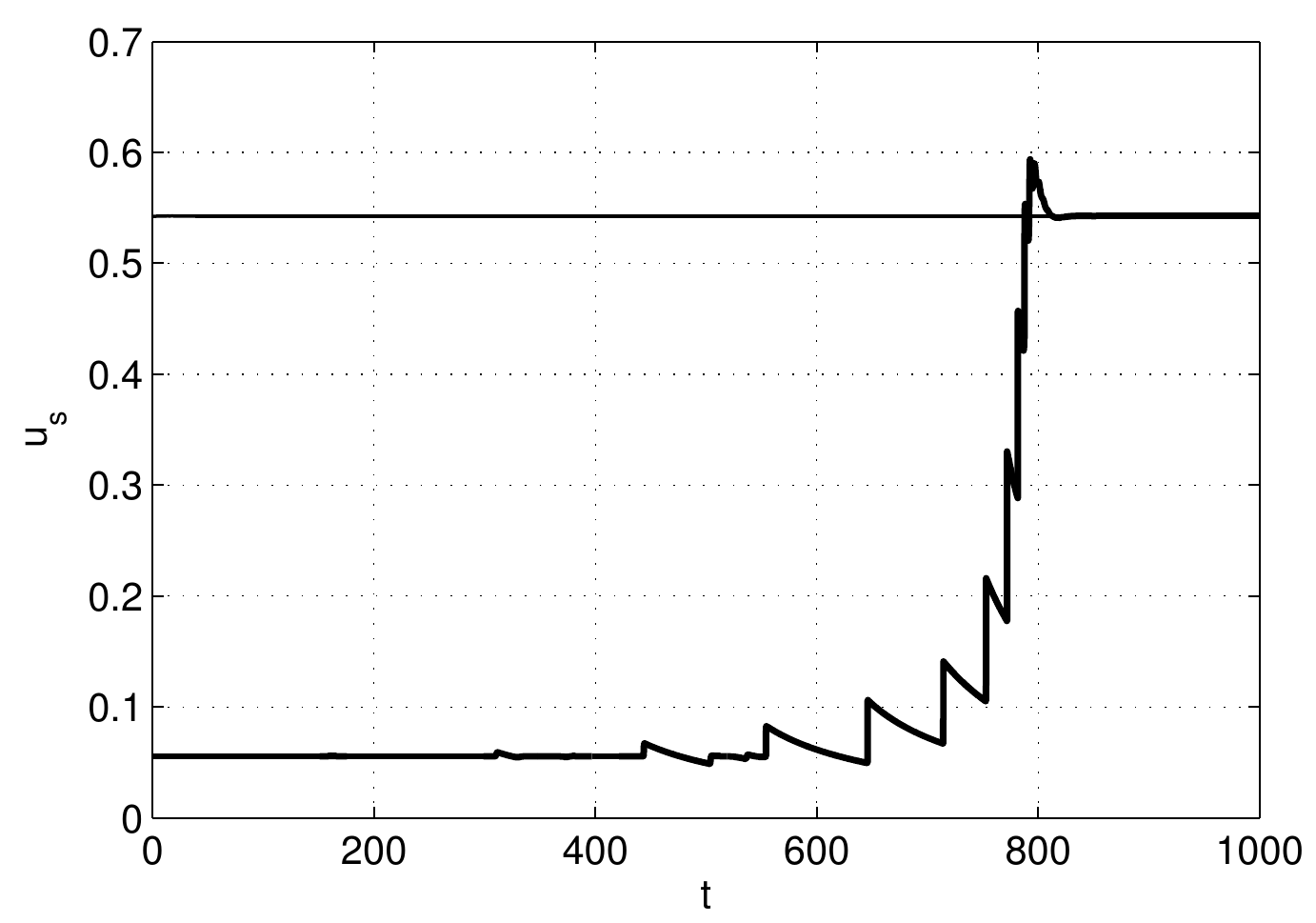}\includegraphics[clip,height=2.5in]{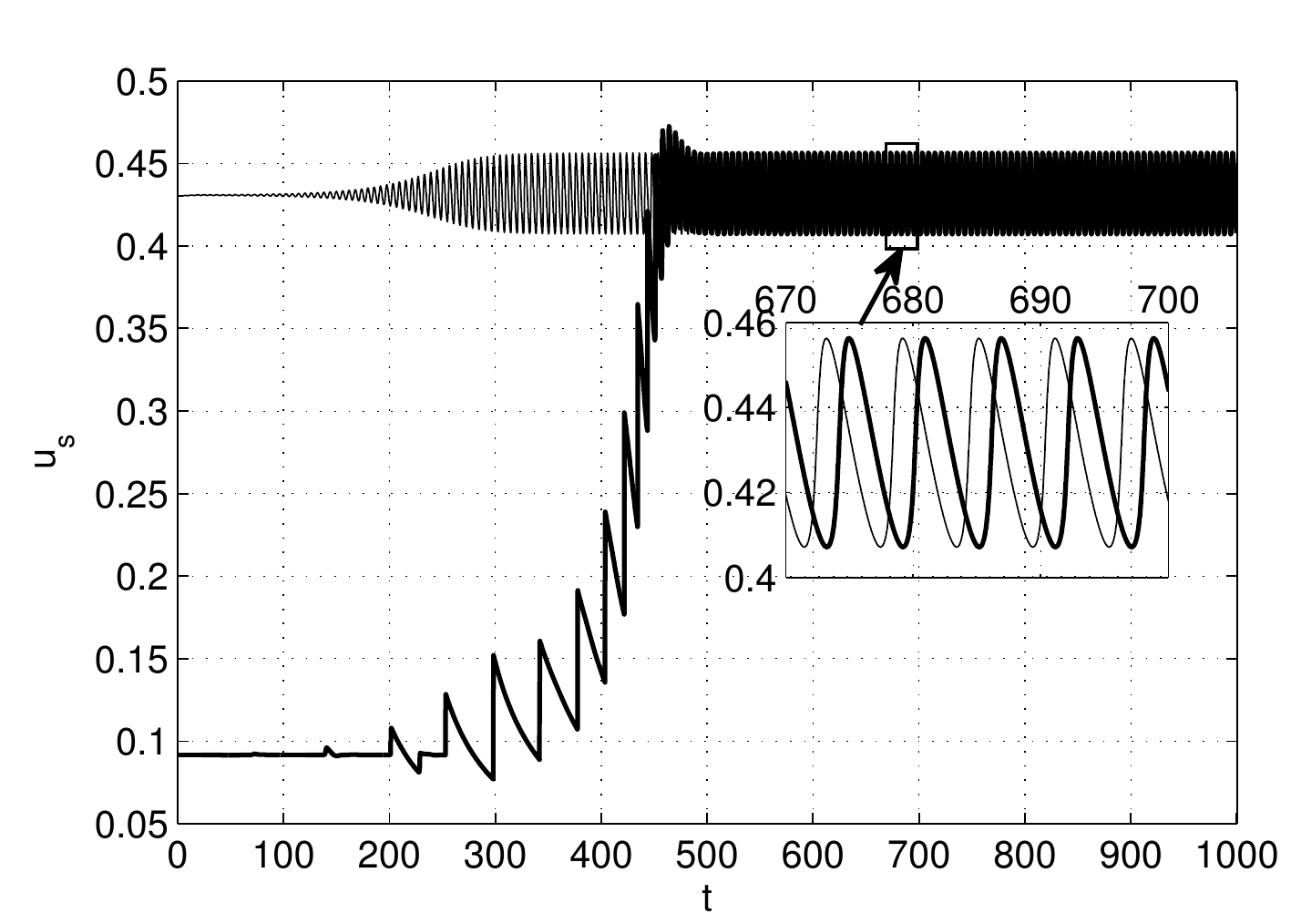}
\par\end{centering}

\caption{\label{fig:Dynamics-with-friction} Time evolution of solutions for
detonation with friction starting with the middle curve of Fig. \ref{fig:Steady-solution-with-friction}
at $\alpha=1$: (a) at $c_{f}=0.1$, the top branch is stable, the
integration is carried out starting both from the top branch (thin
curve) and the bottom branch (thick curve); (b) the same, but at $c_{f}=0.125$,
which corresponds to unstable solutions around the top branch. The
pulsating instability in (b) is due purely to the presence of friction. }
\end{figure}

We find that as we increase $c_{f}$ along the top branch, there is
a critical value of $c_{f}$ above which the detonation becomes unstable,
indicating that the losses have a destabilizing effect. Figure \ref{fig:Dynamics-with-friction}
shows the computed solutions at $c_{f}=0.1$, corresponding to a stable
state on the upper branch, and $c_{f}=0.125$, corresponding to an
unstable state on the upper branch. Note that the instability of the
steady-state solutions on the top branch is associated with a transition
to a limit cycle, likely arising through a Hopf bifurcation when $c_{f}$
exceeds a critical value. These oscillations take place around the
steady-state solution.

As we solve the problem starting on the bottom branch, we find that
the steady-state solution on the branch is indeed unstable, but, unlike
the solutions on the top branch, there is no oscillation around the
bottom branch. The solution tends in fact toward the top branch with
time, indicating that the bottom branch is generally a repelling equilibrium
while the top branch is attracting. The dynamics of this instability
is quite different from that on the top branch. It involves a generation
of internal shock waves in the reaction zone that overtake the lead
shock and, eventually, after multiple such overtakings, the solution
settles on the top branch. The discontinuous behavior of the thick
curves in Fig. \ref{fig:Dynamics-with-friction} occurs precisely
when an internal shock wave catches up with the lead shock. At that
moment, there is a rapid increase of $u_{s}$. The general trend of
the solution appears to be physically reasonable, reflecting the strong
instability of the lower branch of the $D$-$c_{f}$ curve and the
attracting character of the upper branch. It is interesting that very
similar behavior was observed in experiments on initiation of spherical
detonation in hydrocarbon-air mixtures \citep{bull1978initiation}.

Now, we look at spherically expanding detonation solutions. The shock-frame
version of (\ref{eq:KFR-planar}) for a diverging detonation is given
by 
\begin{equation}
\frac{\partial u}{\partial t}+\frac{1}{2}\frac{\partial}{\partial x}\left(u^{2}-uu_{s}\right)=f\left(x,u_{s}\right)-\frac{u^{2}}{x+r_{s}},\label{eq:KFR-spherical}
\end{equation}
where $r_{s}\left(t\right)$ denotes the shock radius such that $dr_{s}/dt=D=u_{s}/2$.
When $\partial u/\partial t$ is dropped, (\ref{eq:KFR-spherical})
can be written as 
\begin{equation}
\frac{du_{0}}{dx}=\frac{f\left(x,u_{s}\right)-\kappa u_{0}^{2}/\left(1+\kappa x\right)}{u_{0}-u_{s}/2},\label{eq:KFR-spherical-steady}
\end{equation}
where $\kappa=1/r_{s}$ is the mean curvature of the shock. This equation
must be solved subject to $u_{0}\left(0\right)=u_{s}$ and to some
appropriate condition at $x=-r_{s}$, i.e., at $r=0$. 

\begin{figure}
\noindent \begin{centering}
\includegraphics[width=3in]{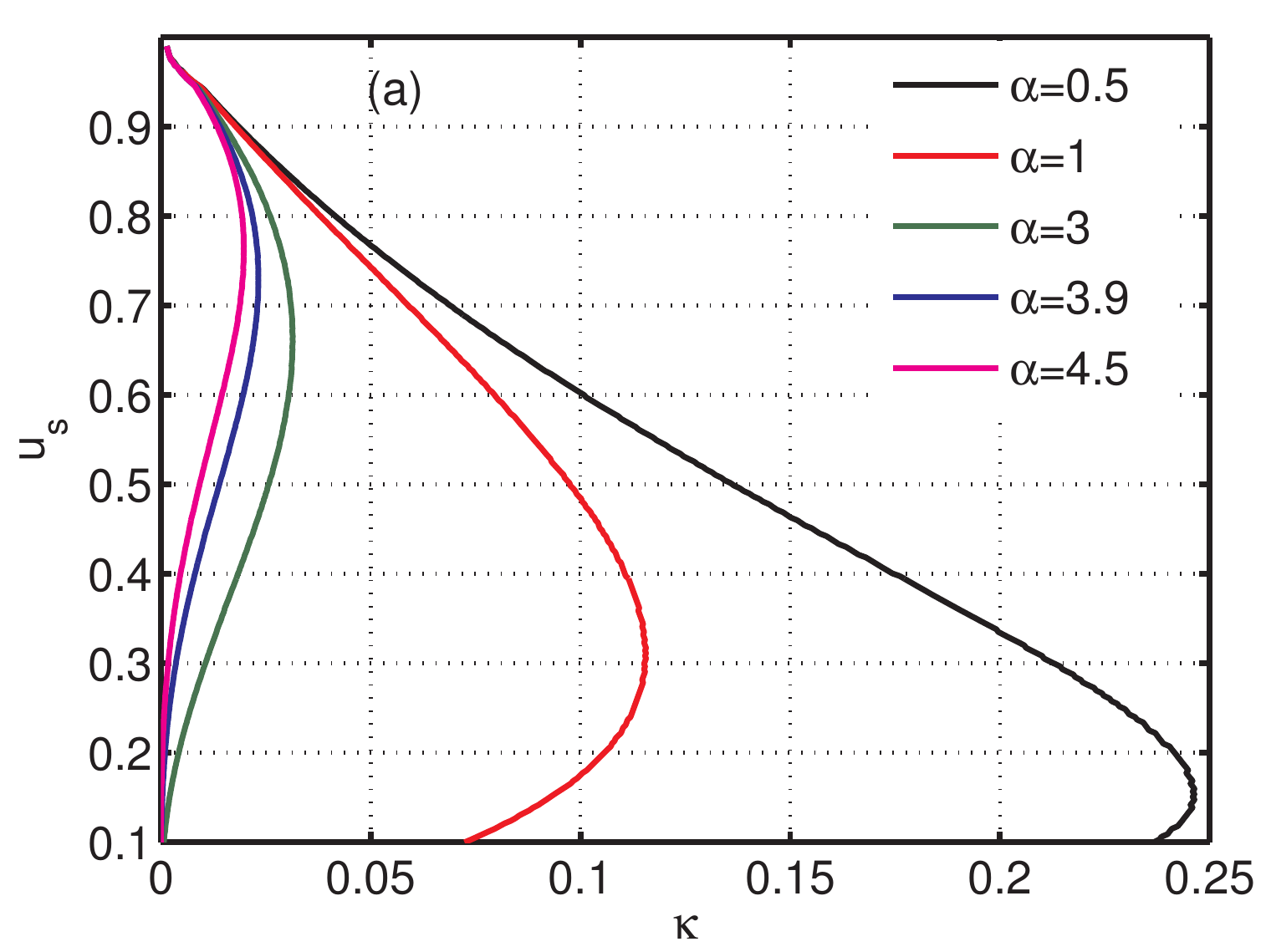}\includegraphics[width=3in]{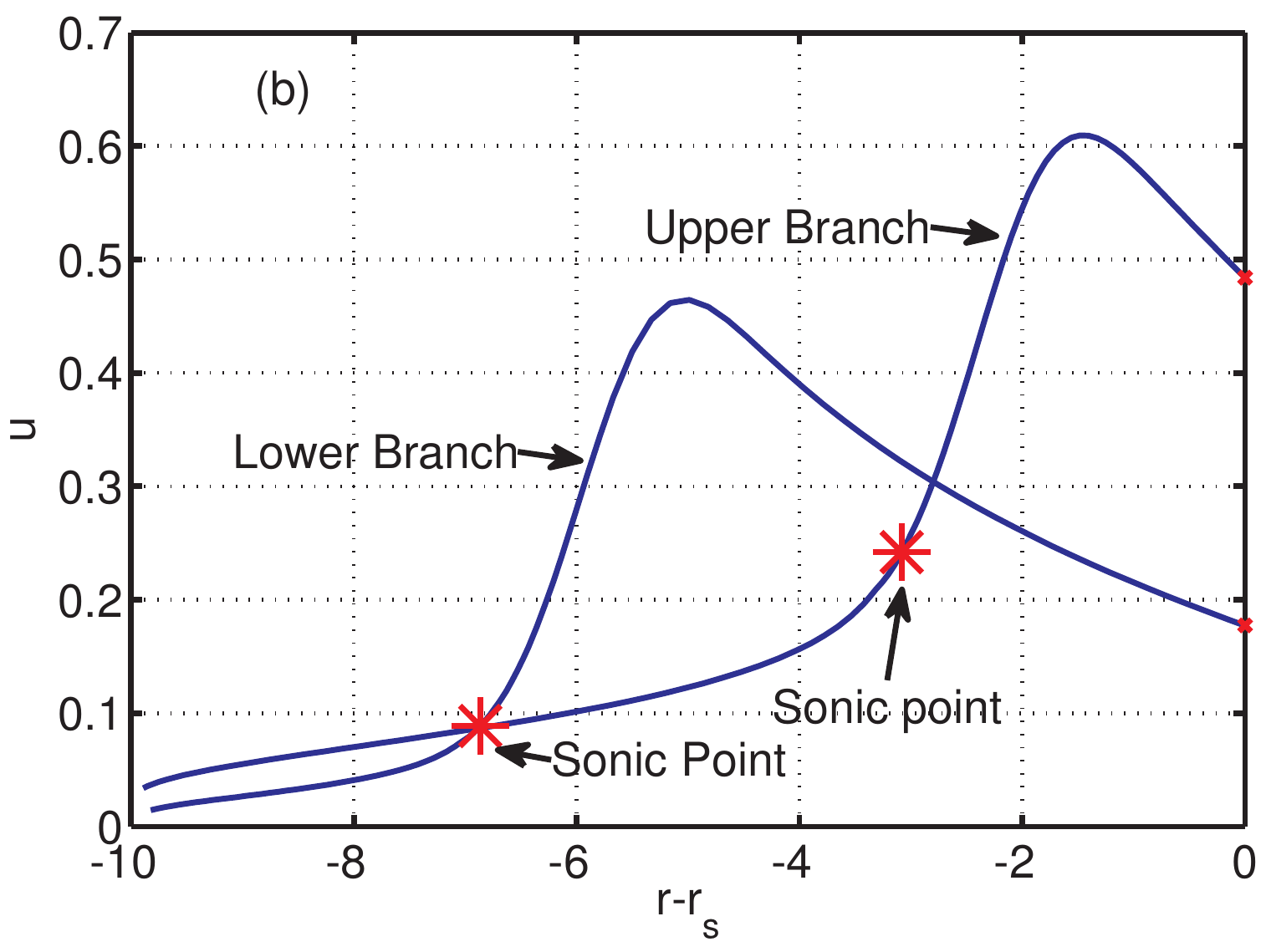}
\par\end{centering}

\caption{\label{fig:Quasi-steady-solution} (a) Quasi-steady $u_{s}-\kappa$
curves at $\beta=0.1$ fixed and variable $\alpha$. (b) The quasi-steady
solution profiles $u_{0}\left(x\right)$ on the top and the bottom
branches of the $u_{s}-\kappa$ curve in (a) at $\alpha=1$ and $\kappa=0.1$.}
\end{figure}

Equation (\ref{eq:KFR-spherical-steady}) is solved using the algorithm
described earlier. In Fig. \ref{fig:Quasi-steady-solution}(a), we
show the computed dependence of $u_{s}$ on $\kappa$ for various
values of $\alpha$ at $\beta=0.1$. The usual turning-point behavior
is seen with the critical curvature decreasing as $\alpha$ increases.
This is similar to that in the Euler detonations wherein the activation
energy leads to the same effects \citep{BdzilStewart07,KasimovStewart05}.
One important difference is that, in Fig. \ref{fig:Quasi-steady-solution}(a),
there are only two branches, the lower branch tending to $u_{s}=0$
and $\kappa=0$, while in the Euler equations, there are in general
three branches, the lower branch tending to $D=c_{a}$, the ambient
sound speed, and $\kappa\to\infty$. In Fig. \ref{fig:Quasi-steady-solution}(b),
we also show the solution profiles that correspond to the $u_{s}-\kappa$
curves in Fig. \ref{fig:Quasi-steady-solution}(a) at a particular
value of $\kappa=0.1$, but at two different values of $u_{s}$, one
on the upper branch and one on the lower. A notable feature of these
profiles is the existence of an internal maximum of $u$, which does
not exist in the planar solution at the same parameters. 

In order to understand better the role of the curvature term in (\ref{eq:KFR-spherical}),
we solve the equation simulating the direct initiation of gaseous
detonation. In the laboratory frame of reference, (\ref{eq:KFR-spherical})
takes the form 
\begin{equation}
\frac{\partial u}{\partial t}+\frac{\partial}{\partial r}\left(\frac{u^{2}}{2}\right)=-\frac{u^{2}}{r}+\begin{cases}
f\left(r-r_{s},u\left(r_{s},t\right)\right), & r<r_{s},\\
0, & r>r_{s}.
\end{cases}\label{eq:KFR-spherical-lab}
\end{equation}
We solve this equation using a fifth-order WENO algorithm \citep{HenrickAslamPowers2006}
and the initial conditions corresponding to a localized source of
the type $u\left(r,0\right)=u_{i}$ at $0<r<r_{i}$ and $u\left(r,0\right)=0$
at $r>r_{i}$. Here, $r_{i}$ is the radius of the initial hot spot
and $u_{i}$ is its ``temperature''. The point-blast initiation
is simulated keeping $r_{i}$ fixed at some small value and varying
$u_{i}$, a measure of the source energy. 
\begin{figure}
\noindent \begin{centering}
\includegraphics[width=3in]{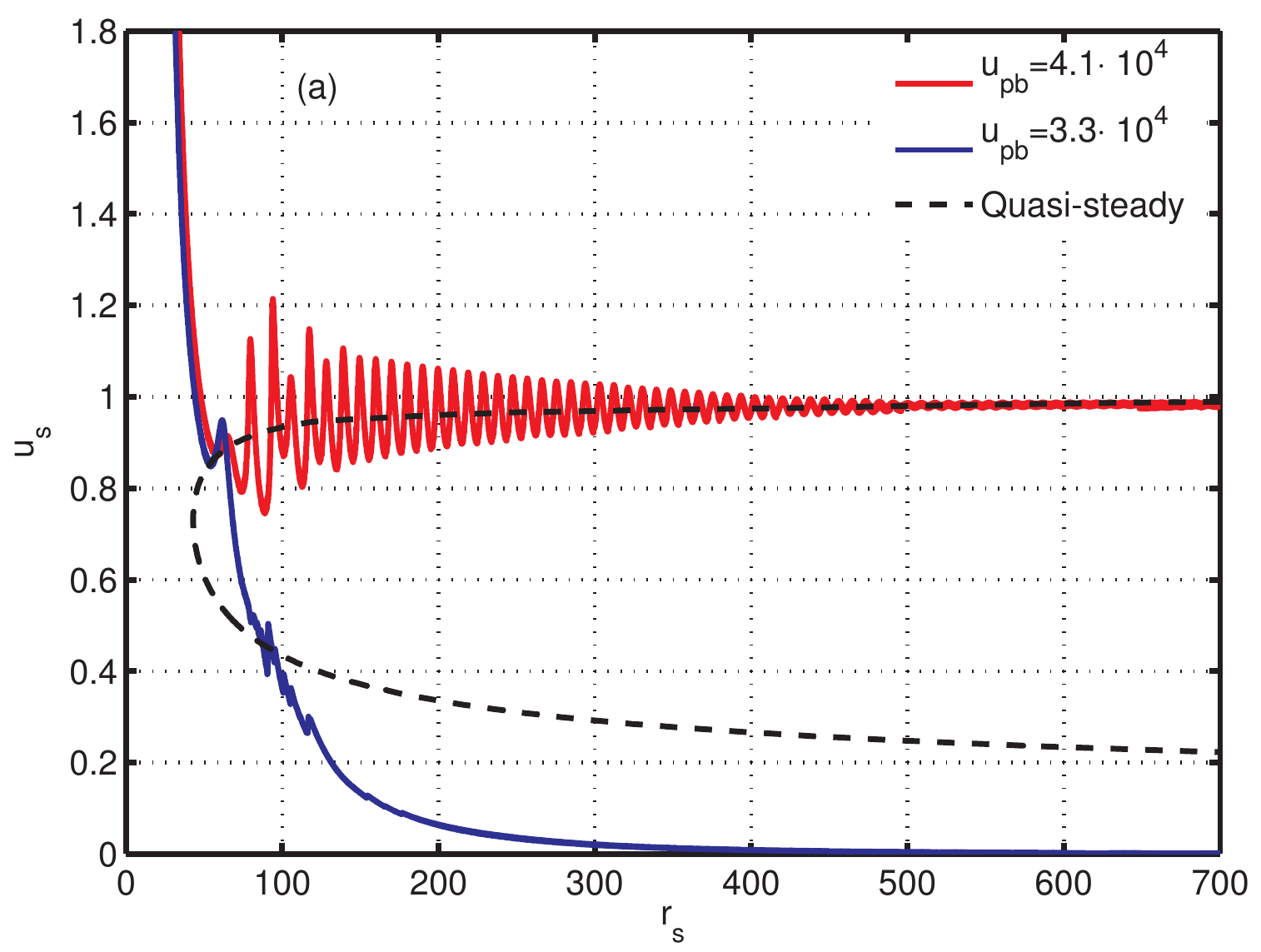}\includegraphics[clip,width=3in]{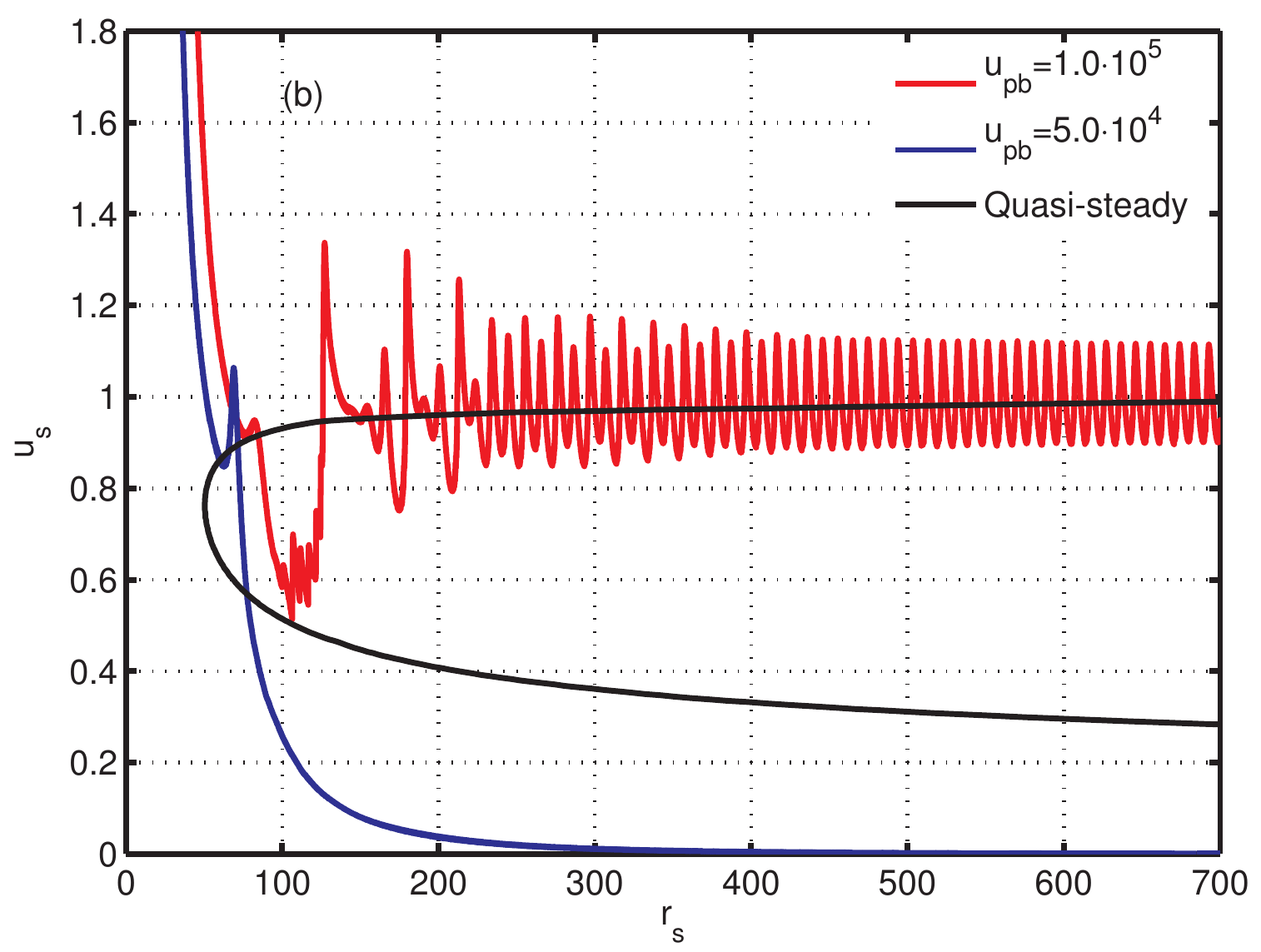}
\par\end{centering}

\caption{\label{fig:Initiation-failure} Initiation and failure of: (a) stable
solutions at $\alpha=3.9$ and (b) unstable solutions at $\alpha=4.5$.
In both figures, $\beta=0.1$, the length of the computational domain
is $L=10^{3}$, and the number of grid points used is $N=10^{4}$.}
\end{figure}

Our findings are displayed in Fig. \ref{fig:Initiation-failure}.
We select two sets of parameters for $\alpha$ and $\beta$ such that
one corresponds to a stable planar solution and the other to an unstable
planar solution. For each case, we vary $u_{i}$ to see if the detonation
initiates or fails. Exactly as in the Euler detonations \citep{watt2005linear},
we observe that above a certain critical value, $u_{ic}$, there is
an initiation; below there is failure. Moreover, the curvature in
our model also plays a destabilizing role. As one can see in Fig.
\ref{fig:Initiation-failure}(a), the detonation that is stable in
the planar case oscillates in the presence of significant curvature.
The oscillations are large in magnitude and irregular at first, around
$r_{s}=100$ to about $r_{s}=150$, before settling down to regular
decaying oscillations. A similar trend is seen in the unstable case,
shown in Fig. \ref{fig:Initiation-failure}(b), where the range of
the irregular oscillations extends from about $r_{s}=120$ to $r_{s}=400$
before settling down to regular periodic oscillations. When the curvature
is significantly diminished, the detonation dynamics is essentially
that of a planar wave. Hence, all the phenomena observed in \citep{kasimov2013model,FariaKasimovRosales-SIAM2014}
carry over to the present study. However, the destabilizing effect
of curvature, clearly seen in Fig. \ref{fig:Initiation-failure},
requires further analysis in order to reveal the underlying mechanisms.
An additional factor that contributes to the instability of the solutions
is $\beta$. For planar solutions, we have shown in \citep{FariaKasimovRosales-SIAM2014}
that smaller $\beta$ leads to more unstable solutions, and we expect
the same effect to be preserved in the curved detonations as well.

\section{Conclusions}

A reactive Burgers equation with nonlocal forcing and appropriate
damping is shown to capture, at a qualitative level, the dynamics
of detonations with friction and of radially diverging detonations.
Using a new integration algorithm, we have found that for curved detonations
and for non-ideal detonations, steady/quasi-steady solutions exist,
which have a characteristic turning-point shape in the plane of the
shock speed versus curvature or a friction coefficient. Unsteady numerical
simulations of our model equation reproduce the dynamics of the point-blast
initiation, capturing the initiation/failure phenomenon. The curvature or the presence of friction are found to play a destabilizing role
in the dynamics of non-ideal detonation. The present calculations
together with our earlier study of the planar model demonstrate that
the reactive Burgers equation is capable of reproducing, qualitatively,
most of the dynamical properties of one-dimensional detonations.\\

\section*{Acknowledgements}
The research reported here was supported by King Abdullah
University of Science and Technology (KAUST).

\bibliographystyle{unsrt}

\appendix

\section{Transonic integration of reactive Euler equations}

Here, we describe an algorithm for numerical integration of the system
of ordinary differential equations (ODE) for the transonic structure
of traveling-wave solutions of reactive Euler equations in one spatial
dimension. Because the algorithm works for a general one-dimensional
system of hyperbolic balance laws, we begin with such a system. Then
we specialize to a system of reactive Euler equations and provide
an example of a weakly curved detonation.

Consider a system of hyperbolic balance laws,
\begin{equation}
\boldsymbol{q}_{t}+\mathbf{F}(\boldsymbol{q})_{x}=\mathbf{s}\left(\mathbf{q}\right),\label{eq:general-balance-law}
\end{equation}
where $\mathbf{q}$ is the vector of unknowns, $\mathbf{F}$ is the
flux vector, and $\mathbf{s}$ is a source term. We look for traveling
wave solutions $\boldsymbol{q}=\boldsymbol{q}(x-Dt)=\boldsymbol{q}(\eta),$
consisting of a shock followed by a smooth flow downstream. The state
upstream of the shock, $\eta>0$, is assumed to be uniform and steady,
$\boldsymbol{q}=\boldsymbol{q}_{a}=constant$. Then, $\mathbf{q}$,
solves 
\begin{equation}
\left(\mathbf{F}(\boldsymbol{q})-D\boldsymbol{q}\right)_{\eta}=\mathbf{s}\label{eq:ode-TWS-general}
\end{equation}
in smooth parts of the flow, where $\eta=0$ is the shock position
and $\eta<0$ is the downstream region. At $\eta=0$, the following
shock conditions are satisfied: 
\begin{equation}
-D\left[\mathbf{q}\right]+\left[\mathbf{F}\right]=0,\label{eq:RH-general}
\end{equation}
with $\left[Z\right]=Z^{+}-Z^{-}$ denoting the jump in the quantity
$Z$ across the shock. The solution of (\ref{eq:RH-general}) can
be written as $\boldsymbol{q}(0_{-})=\boldsymbol{q}_{RH}(D,\boldsymbol{q}_{a})$.
The shock speed, $D$, is an unknown of the problem and must be found
together with the profiles of $\mathbf{q}$ at $\eta<0$.

A well-known difficulty in solving (\ref{eq:ode-TWS-general}) arises
when one of the eigenvalues of the matrix, $\mathbf{F}_{\mathbf{q}}-D\mathbf{I}$,
where $\mathbf{F}_{\mathbf{q}}\equiv\partial\mathbf{F}/\partial\boldsymbol{q}$,
vanishes at some point $\eta_{*}<0$ (a sonic point), thus producing
a singular system of ODE, $\left(\mathbf{F}_{\mathbf{q}}-D\mathbf{I}\right)\mathbf{u}_{\eta}=\mathbf{s}$
\citep{higgins2012steady}. This feature is an essential ingredient
of any self-sustained shock wave and is thus relevant in many applications
where such traveling shock-wave solutions arise (e.g., traffic flow
problems \citep{FlynnKNRS-PRE09}, hydraulic jumps \citep{kasimov2008stationary}).
Should there be a vanishing eigenvalue, a regularity condition is
called upon where, for boundedness of $\boldsymbol{q}_{\eta}$, it
is required that 
\begin{equation}
\mathbf{l}_{*}\cdot\mathbf{s}_{*}=0\quad\text{when}\quad\lambda_{*}=0,\label{eq:regularity-cond-general}
\end{equation}
where $\lambda_{*}$ is the special eigenvalue of $\mathbf{F}_{\mathbf{q}}-D\mathbf{I}$
that vanishes at $\eta_{*}$ and $\mathbf{l}_{*}$ is the corresponding
left eigenvector. Condition (\ref{eq:regularity-cond-general}) serves
as a closure condition that identifies admissible shock speeds, $D$. 

Because analytic integration of (\ref{eq:ode-TWS-general}) is rarely
possible, a numerical procedure is required. When a vanishing eigenvalue
exists somewhere in the flow, we need a numerical algorithm to determine
the values of $D$ for which (\ref{eq:regularity-cond-general}) is
satisfied. Importantly, the location of the critical point is unknown
\emph{a priori}. A simple approach to solving this problem is to make
a guess for $D$ and integrate from $\eta=0$ up to the singular point,
and then check whether or not $\mathbf{l}_{*}\cdot\mathbf{s}_{*}=0$
is satisfied. This is a numerically ill-conditioned procedure since
the system becomes stiffer as one approaches the singular point, the
latter having a saddle-point nature.

Our integration procedure avoids the numerical problems associated
with the presence of a sonic point. The key idea is based on the use
of a new dependent variable given by 
\begin{equation}
\boldsymbol{z}=\mathbf{G}(\boldsymbol{q};D)=\mathbf{F}\left(\boldsymbol{q}\right)-D\boldsymbol{q}.\label{eq:z_def}
\end{equation}
The governing system of ODEs written in terms of $\mathbf{z}$ becomes
\begin{equation}
\boldsymbol{z}_{\eta}=\mathbf{s}(\boldsymbol{q}),\label{eq:flux-var-ode-general}
\end{equation}
and needs to be solved subject to the shock conditions, $\boldsymbol{z}(0)=\mathbf{F}\left(\boldsymbol{q}_{0}\right)-D\boldsymbol{q}_{0},$
with $\mathbf{q}_{0}$ denoting the post-shock state. In order for
this change of variables to be successful, it must be invertible so
that $\boldsymbol{q}=\mathbf{G}^{-1}(\boldsymbol{z},D)$. The inversion
is guaranteed to be well defined as long as the Jacobian, $\mathbf{G}_{\mathbf{q}}=\mathbf{F}_{\mathbf{q}}-D\mathbf{I}$,
is not singular, which is the case away from sonic points. It is important
to note that, in general, the inversion results in multiple solution
branches. In order to choose the correct branch, we need to ensure
that $\mathbf{G}^{-1}\left(\boldsymbol{z}(0)\right)=\boldsymbol{q}_{0}$. 

The main advantage of the new variable is that (\ref{eq:flux-var-ode-general})
is not stiff even as one approaches the singular point and thus the
problem of finding the values of $D$ such that (\ref{eq:regularity-cond-general})
is satisfied becomes regular. The analytical inversion of $\mathbf{G}$
may not in general be possible as it depends on the specific form
of the equation of state. Nevertheless, the general procedure remains
valid and, once the sonic points are found, the inversion can be done
numerically.

To specialize the previous analysis to one-dimensional reactive Euler
equations, we begin with the equations written in conservation form:
\begin{eqnarray}
\rho_{t}+\left(\rho u\right)_{x} & = & s_{1},\\
\left(\rho u\right)_{t}+\left(\rho u^{2}+p\right)_{x} & = & s_{2},\\
\left(\rho e\right)_{t}+\left(\rho ue+pu\right)_{x} & = & s_{3},\\
\left(\rho\lambda\right)_{t}+\left(\rho u\lambda\right)_{x} & = & s_{4}.
\end{eqnarray}
We have chosen to keep $\mathbf{s}$ general for now. It can account
for such effects as curvature, heat and momentum losses, area changes,
etc. For simplicity, we assume a perfect gas equation of state and
therefore $e=pv/\left(\gamma-1\right)-Q\lambda+u^{2}/2$, $p=\rho RT$,
where $Q$ is the heat of reaction, $\lambda$ is the heat-release
progress variable, and $R$ is the universal gas constant. Now, let
$q_{1}=\rho$, $q_{2}=\rho u$, $q_{3}=\rho e$, $q_{4}=\rho\lambda$.
Then, $p=\left(\gamma-1\right)\left(q_{3}-\frac{q_{2}^{2}}{2q_{1}}+Qq_{4}\right)$.
In terms of these conserved quantities, we find that 
\begin{equation}
\mathbf{F}\left(\boldsymbol{q}\right)=\begin{pmatrix}q_{2}\\
q_{2}^{2}/q_{1}+\left(\gamma-1\right)\left(q_{3}-\frac{q_{2}^{2}}{2q_{1}}-Qq_{4}\right)\\
q_{2}q_{3}/q_{1}+\frac{q_{2}}{q_{1}}\left(\gamma-1\right)\left(q_{3}-\frac{q_{2}^{2}}{2q_{1}}-Qq_{4}\right)\\
\frac{q_{2}}{q_{1}}q_{4}
\end{pmatrix}
\end{equation}
and the eigenvalues of $\mathbf{F}_{\mathbf{q}}$ (which we do not
write for brevity) give the well-known characteristic speeds of the
Euler equations in the shock-attached frame:
\begin{eqnarray*}
\lambda_{1} & = & q_{2}/q_{1}-\sqrt{\frac{-q_{1}^{2}\left(q_{2}^{2}-2q_{1}q_{3}+2Qq_{1}q_{4}\right)(\gamma-1)\gamma}{2q_{1}^{4}}}=u-c-D,\\
\lambda_{2} & = & q_{2}/q_{1}=u-D,\\
\lambda_{3} & = & q_{2}/q_{1}=u-D,\\
\lambda_{4} & = & q_{2}/q_{1}+\sqrt{\frac{-q_{1}^{2}\left(q_{2}^{2}-2q_{1}q_{3}+2Qq_{1}q_{4}\right)(\gamma-1)\gamma}{2q_{1}^{4}}}=u+c-D,
\end{eqnarray*}
where the sound speed is given by $c=\sqrt{\frac{-\left(q_{2}^{2}-2q_{1}q_{3}+2Qq_{1}q_{4}\right)(\gamma-1)\gamma}{2q_{1}^{2}}}=\sqrt{\frac{\gamma p}{\rho}}$.
In order to obtain the regularity condition at the sonic point, we
need to know the left eigenvector associated with the forward characteristic.
It is given by 
\begin{eqnarray*}
l_{4} & = & \begin{pmatrix}u^{2}\left(\gamma-1\right)-uc, & c-u\left(\gamma-1\right), & \left(\gamma-1\right), & Q\left(\gamma-1\right)\end{pmatrix}.
\end{eqnarray*}
Thus, should there be a sonic point in the flow $\left(\lambda_{4}=0\right)$,
it is necessary that $\mathbf{l}_{4}\cdot\mathbf{s}$ should vanish
at the sonic point in order for $\boldsymbol{q}_{\eta}$ to be bounded. 

Following the general procedure outlined above, we define
\begin{eqnarray}
z_{1} & = & q_{2}-Dq_{1},\nonumber \\
z_{2} & = & q_{2}^{2}/q_{1}+\left(\gamma-1\right)\left(q_{3}-\frac{q_{2}^{2}}{2q_{1}}-Qq_{4}\right)-Dq_{2},\label{eq:transformation-cons-to-flux}\\
z_{3} & = & q_{2}q_{3}/q_{1}+\frac{q_{2}}{q_{1}}\left(\gamma-1\right)\left(q_{3}-\frac{q_{2}^{2}}{2q_{1}}-Qq_{4}\right)-Dq_{3},\nonumber \\
z_{4} & = & \frac{q_{2}}{q_{1}}q_{4}-Dq_{4}.\nonumber 
\end{eqnarray}
We obtain the inverse, $\boldsymbol{q}=\boldsymbol{q}(z_{1},z_{2},z_{3},z_{4})$,
as 
\begin{eqnarray}
q_{2} & = & z_{1}+Dq_{1}\nonumber \\
q_{3} & = & \frac{D^{2}q_{1}^{2}z_{1}+2Dq_{1}z_{1}^{2}+z_{1}^{3}-Dq_{1}^{2}z_{2}-q_{1}z_{1}z_{2}+q_{1}^{2}z_{3}}{q_{1}z_{1}}\nonumber \\
q_{4} & = & \frac{q_{1}z_{4}}{z_{1}}\label{eq:cons-to-flux}
\end{eqnarray}
with 
\begin{equation}
q_{1}=\frac{\gamma z_{1}(z_{2}-Dz_{1})\pm\sqrt{z_{1}^{2}\left(D^{2}z_{1}^{2}-2Dz_{1}z_{2}+\gamma^{2}z_{2}^{2}-2\left(\gamma^{2}-1\right)z_{1}(z_{3}+Qz_{4})\right)}}{D^{2}z_{1}-2Dz_{2}+2(\gamma-1)(z_{3}+Qz_{4})}.\label{eq:q_1}
\end{equation}
The choice of the inversion branch depends on which branch of the
square root in (\ref{eq:q_1}) is chosen. We note that the expression
under the square root is 
\begin{eqnarray}
\delta & \equiv & z_{1}^{2}\left(D^{2}z_{1}^{2}-2Dz_{1}z_{2}+\gamma^{2}z_{2}^{2}-2z_{1}(z_{3}+Qz_{4})\left(\gamma^{2}-1\right)\right)=\rho^{4}\lambda_{1}^{2}\lambda_{3}^{2}\lambda_{4}^{2},
\end{eqnarray}
i.e., it is a perfect square that vanishes only when one of the eigenvalues
of the Jacobian, $\mathbf{F}_{\mathbf{q}}-D\mathbf{I}$, becomes zero.
One can thus simplify $q_{1}$ as 
\begin{eqnarray}
q_{1} & = & \frac{\rho\left(\gamma(u-D)^{2}+c^{2}\right)-\text{sign}(u-D)\rho\left|\left(u-D+c\right)\left(u-D-c\right)\right|}{(\gamma-1)\left((u-D)^{2}+\frac{2}{\gamma-1}c^{2}\right)}.
\end{eqnarray}
The correct branch of the transformation is selected by requiring
that $\boldsymbol{q}(\boldsymbol{z}(0))=\boldsymbol{q}(0)$, which
can be seen to be the negative branch\@. Across the sonic point,
the solution branch changes.

In order to illustrate the previous calculation with a well-known
example, we consider the small curvature approximation of the reactive
Euler equations, which can be written in a shock-attached frame as
(see, e.g., \citep{klein1993relation}), 
\begin{eqnarray}
\left(\rho\left(u-D\right)\right)_{\eta} & = & -\kappa\rho u,\\
\left(p+\rho u\left(u-D\right)\right)_{\eta} & = & -\kappa\rho u^{2},\\
\left(\rho\left(u-D\right)e+pu\right)_{\eta} & = & -\kappa\left(\rho ue+pu\right),\\
\left(\rho(u-D)\lambda\right)_{\eta} & = & \rho\omega-\kappa\rho u\lambda.
\end{eqnarray}
Here, $\omega=\omega\left(p,\rho,\lambda\right)$ is a general rate
function, not necessarily of Arrhenius form. As before, we define
new dependent variables as in (\ref{eq:transformation-cons-to-flux})
and the inverse as in (\ref{eq:cons-to-flux}). The system written
in terms of the new variable is simply $\boldsymbol{z}_{\eta}=\mathbf{s}(\boldsymbol{z})$.
In this particular case, the total enthalpy, $H=\frac{\gamma p}{\left(\gamma-1\right)\rho}+\frac{\left(u-D\right)^{2}}{2}-\lambda Q$,
can be shown to be a conserved quantity. Therefore, using the upstream
state to rescale the variables with respect to $p_{a}$, $\rho_{a}$,
and $\sqrt{p_{a}/\rho_{a}}$, we find that $H=\frac{D^{2}z_{1}-2Dz_{2}+2z_{3}}{2z_{1}}=H_{0}=\frac{\gamma}{\gamma-1}+\frac{D^{2}}{2}.$
Then, we eliminate $z_{3}$ in favor of the remaining variables, $z_{3}=\frac{\gamma}{\gamma-1}z_{1}+Dz_{2},$
to arrive at the following system: 
\begin{eqnarray}
\left(z_{1}\right)_{\eta} & = & -\kappa q_{2},\\
\left(z_{2}\right)_{\eta} & = & -\kappa\frac{q_{2}^{2}}{q_{1}},\\
\left(z_{4}\right)_{\eta} & = & \frac{-\kappa q_{2}q_{4}+\omega}{q_{1}},
\end{eqnarray}
which is free from singularity. It should be integrated numerically
from the shock to the sonic point with the negative branch in (\ref{eq:q_1})
and, if necessary, further from the sonic point using the positive
branch.
\end{document}